\documentclass{webofc}
\usepackage[varg]{txfonts} 
\usepackage{subfigure}
\usepackage{float}

\usepackage{xcolor}

\begin{document}

\title{Relativistic Dissipative Magnetohydrodynamics for accretion disks}

\author{\firstname{Khwahish} \lastname{Kushwah}\inst{1}\fnsep\thanks{\email{khwahish_kushwah@id.uff.br}} \and
        \firstname{Gabriel} \lastname{S.~Denicol}\inst{1}\fnsep\thanks{\email{gsdenicol@id.uff.br}}
        }

\institute{Instituto de F\'isica, Universidade Federal Fluminense (UFF), Niter\'oi,
24210-346, RJ, Brazil
          }

\abstract{
We derive a relativistic magnetohydrodynamics (RMHD) theory for a dilute electron–ion gas governed by the Boltzmann–Vlasov equation, using the method of moments. This yields an extended MHD framework beyond standard astrophysical formulations, which typically include only the shear-stress component parallel to the magnetic field. We analyze our framework in the linear regime and show that it leads to the firehose instability when the background longitudinal pressure becomes large. In these extreme scenario, the transverse and semi-transverse shear-stress components become large and may play a role in accretion disk dynamics.
}

\maketitle
\section{Introduction}
\label{intro}
Relativistic magnetohydrodynamics (RMHD) provides a theoretical framework for understanding the dynamics of relativistic fluids in the presence of strong magnetic fields. This framework is pivotal in various astrophysical contexts as well as in the high-energy heavy-ion collisions \cite{Armas_2022}. During the early stages of these collisions, intense magnetic fields are generated, reaching peak values of $10^{19}$ Gauss at the Relativistic Heavy Ion Collider (RHIC) and $10^{20}$ Gauss at the Large Hadron Collider (LHC). Similarly, in astrophysical scenarios, such as around magnetars and accretion disks, magnetic fields can reach strengths as high as $10^{15}$ Gauss \cite{Hattori_2022}. 

Theories derived to model the dynamics of accretion disks in the presence of strong magnetic fields assume that the longitudinal component of the shear stress tensor (with respect to the magnetic field's direction)  becomes dominant, with the transverse components being considered negligible \cite{Chandra_2015, Foucart_2015, Most_2022, Cordeiro_2024}. On the other hand, for simulations of heavy-ion collision, recent calculations have shown that the transverse components of the shear-stress tensor are as important as the longitudinal ones and may play a role in the evolution of the plasma \cite{Kushwah_2024}. Motivated by this distinction, we derive the equations of motion for the shear-stress tensor in the context of accretion disks, modeling the system as a binary mixture of ions and electrons. In doing so, we consider all components of the shear stress tensor with respect to the magnetic field, and explore the validity of the "longitudinal-only" assumption. By examining the contributions of the transverse components, we assess whether they play a significant role in the system dynamics.

\section{Equations of motion}
\label{Equations of motion}
The evolution dynamics of the hydrodynamical system is governed by the conservation of energy-momentum tensor, $\displaystyle{\partial_{\mu} T^{\mu \nu}  = 0.}$
 For plasmas in the presence of strong magnetic fields, $T^{\mu\nu}$ can be written as
 \cite{Denicol_resistive, Denicol_2018}
\begin{equation}
     T^{\mu \nu}= \left(\epsilon + \frac{B^2}{2}\right) u^{\mu} u^{\nu}-\Delta^{\mu \nu}\left(P+\frac{B^2}{2}\right) - B^{\mu} B^{\nu} +\pi^{\mu \nu} , 
\end{equation}
where $\epsilon$ is the energy density, $u^{\mu}$ is the fluid four-velocity, $P$ is the thermodynamic pressure, $\pi^{\mu \nu}$ is the total shear-stress tensor, $B^\mu$ is magnetic field four-vector such that $B^2 \equiv -B^\mu B_\mu$. We have used the notation $\displaystyle{A^{\langle\mu\nu\rangle} \equiv \Delta^{\mu\nu}_{\alpha\beta}A^{\alpha\beta}}$, with $\displaystyle{\Delta^{\mu\nu}_{\alpha\beta}  \equiv \left(\Delta^{\mu}_\alpha \Delta^{\nu}_\beta + \Delta^{\nu}_\alpha\Delta^{\mu}_\beta - 2/3\Delta^{\mu\nu}\Delta_{\alpha\beta}\right)/2}$, $\displaystyle{\Delta^{\mu\nu} = g^{\mu\nu} - u^{\mu}u^{\nu}}$, and $\displaystyle{ g^{\mu\nu} = diag\ (1,-1,-1,-1)}$.

\subsection{Boltzmann equation}
\label{boltzmann equation}
We consider a relativistic locally electrically neutral fluid composed of two classical particle species namely ions with charge $+q$ and electrons with charges $-q$. Both species are assumed to have vanishing dipole moment and spin. The state of each species is described by its single-particle distribution function, $f^\pm$, where the upper sign $(+)$ refers to ions and the lower sign $(-)$ to electrons. Their charges are denoted as $q^\pm = \pm q$. 
We also assume constant electric charge chemical potential, $\mu_q$, and baryon chemical potential, $\mu_B$. The dynamics of the distribution functions is governed by the Boltzmann-Vlasov equation,
\begin{equation}
\setlength\abovedisplayskip{3pt}
\setlength\belowdisplayskip{3pt}
        k^{\mu} \partial_\mu f^{\pm}_{k} + q^{\pm} k_{\nu} F^{\mu \nu} \frac{\partial}{\partial k^{\mu}}f_k^{\pm}  = C[f^{\pm},f^\mp],
\end{equation}
where we consider only binary elastic collisions\footnote{For the explicit form of the collision term, see \cite{Kushwah_2024}.}. The electromagnetic field contribution enters through the Faraday tensor $F^{\mu\nu}$, which in the absence of an electric field, can be written as $\displaystyle{F^ {\mu\nu} \equiv \epsilon^{\mu\nu\alpha\beta}u_\alpha B_\beta.}$

\subsection{Exact equations of motion}
\label{sub:exact eom}
For a fluid consisting of two particle species, we directly calculate the time derivative of the total shear stress tensor, $\pi^{\mu\nu} = \pi_+^{\mu\nu} + \pi_-^{\mu\nu}$, and the relative shear stress tensor of the system, $\delta\pi^{\mu\nu} \equiv \pi_+^{\mu\nu} - \pi_-^{\mu\nu}$, where $\pi^{\mu\nu}_\pm$ denotes the individual shear stress tensor for each particle species. We employ the 14-moment approximation, along with the assumption that ions are massive, $\displaystyle{z \equiv m_p / T \gg 1}$, and electrons are ultra-relativistic, $\displaystyle{m_e / T \sim 0}$, to compute the collision term \cite{DNMR}. Here, $m_p$ is the mass of the ion, $m_e$ is the mass of the electron, and $T$ is the local temperature of the plasma. The resulting equations of motion are:
\begin{multline}
    \dot{\pi}^{\langle\mu\nu\rangle} + \Sigma_\pi \pi^{\mu\nu} + \hat{\Sigma}_\pi \delta \pi^{\mu\nu} - |\mathbf{q}| J_\omega B b^{\lambda\langle\mu} \pi^{\nu\rangle}_{\lambda}  + |\mathbf{q}|  \hat{J}_\omega Bb^{\lambda\langle\mu}  \delta\pi^{\nu\rangle}_{\lambda}  \\  =   \frac{18}{5}n_e T  \sigma^{\mu \nu}  - \frac{10}{7} \sigma_\lambda^{\langle\mu} 
    \pi^{\nu\rangle\lambda}  - \frac{2}{7} \sigma_\lambda^{\langle\nu}  \delta\pi^{\mu\rangle\lambda}  
    +\frac{1}{2}\pi^{\lambda\langle\mu} \omega^{\nu\rangle} _{\,\, \lambda} - \frac{4\theta}{3} \pi^{\mu\nu}  - \frac{\theta}{6} \delta\pi^{\mu\nu}, 
\\
    \delta\dot{\pi}^{\langle \mu \nu \rangle} + \Sigma_{\delta\pi} \delta\pi^{\mu\nu} +  
    \hat{\Sigma}_{\delta\pi}\pi^{\mu\nu} - |\mathbf{q}|  J_\omega B b^{\lambda\langle\mu} \delta\pi^{\nu\rangle}_{\lambda} + |\mathbf{q}|   \hat{J}_\omega 
    Bb^{\lambda\langle\mu}\pi^{\nu\rangle}_{\lambda} \hspace{4.5cm}\\ =  \frac{2}{5}n_e T \sigma^{\mu \nu}   - \frac{10}{7}  \sigma_\lambda^{\langle\nu} 
    \delta\pi^{\mu\rangle\lambda}  - \frac{2}{7}   \sigma_\lambda^{\langle\nu} \pi^{\mu\rangle\lambda} +\frac{1}{2}\delta\pi^{\lambda\langle\mu} \omega^{\nu\rangle} _{\,\,\lambda} - \frac{4\theta}{6} \delta\pi^{\mu\nu} - \frac{\theta}{6} \pi^{\mu\nu}. \hspace{0.6cm}  
    \end{multline}
    \color{black}
Above, we defined the electron charge density, $\displaystyle{n_e}$, the shear tensor, $\displaystyle{\sigma^{\mu\nu} \equiv \nabla^{\langle\mu} \ u^{\nu\rangle}}$, the expansion scalar, $\displaystyle{\theta \equiv \nabla_\mu u^\mu}$ and the vorticity tensor, $\displaystyle{\omega^{\mu \nu} = (\nabla^{\mu}u^\nu - \nabla^\nu u^\mu)/2}$, with $\nabla_\mu = \Delta^\nu_\mu\partial_\nu$ being the spatial projected gradient. The brackets denotes the double, symmetric and traceless projection of the tensor orthogonal to $u^\mu$. In addition, $b^{\mu\nu}$ = $\epsilon^{\mu\nu\alpha\beta} u_\alpha b_\beta$ is a dimensionless antisymmetric tensor where $b^\mu \equiv$ $B^\mu / B$ specifies the direction of the magnetic field. Further, we have defined the following transport coefficients:
\small\begin{align*}
\Sigma_{\pi} & = \frac{n_{e}}{2}\biggl(\frac{3}{5}\,\sigma_T^{--} + 2\sigma_{T}^{+-} 
+ \frac{16}{5\sqrt{z\pi}}\,\sigma_{T}^{++}\biggr)\, ,\qquad
\hat{\Sigma}_{\pi}  = \frac{n_{e}}{2}\biggl(\frac{16}{5\sqrt{z\pi}}\,\sigma_{T}^{++} 
+ 16\frac{\sigma_{T}^{+-}}{z}
- \frac{3}{5}\,\sigma_T^{--}
- 2\,\sigma_{T}^{+-}\biggr),\\
\Sigma_{\delta\pi} & = \frac{n_{e}}{2}\biggl(\frac{3}{5}\,\sigma_T^{--} 
- 2\,\sigma_{T}^{+-}
+ 32 \frac{\sigma_{T}^{+-}}{z}\, 
+ \frac{16}{5\sqrt{\pi\,z}}\,\sigma_{T}^{++}\biggr)\, ,\quad 
\hat{\Sigma}_{\delta\pi} = \frac{n_{e}}{2}\biggl(-\frac{3}{5}\,\sigma_T^{--}
- 2\,\sigma_{T}^{+-}
+ 16\,\frac{\sigma_{T}^{+-}}{z}
+ \frac{16}{5\sqrt{\pi\,z}}\,\sigma_{T}^{++}\biggr).
\end{align*}\normalsize
Here $\sigma_T^{ij}$ denotes the cross section for interactions between species $i$ and $j$, with $ i,j \in \{+,-\}$. In the limit of asymptotically large $z$, we can also approximate $J_\omega \approx \beta_0\left( 1-\frac{5}{z}\right) \text{\, and \,}\hat{J}_\omega \approx \beta_0\left( 1+\frac{5}{z}\right)$.

\section{Linear Stability and the Firehose Instability}
\label{lineari stability analysis}
To assess the stability properties of the system in the presence of a significant longitudinal pressure, we perform the following linear stability analysis around a dissipative state. We consider the following perturbations,  hereafter denoted as $\Delta$,
\begin{equation}
    \begin{split}
        \varepsilon=\varepsilon_{0}+\Delta \varepsilon,\qquad n=\Delta n, \qquad u^{\mu}=u_{0}^{\mu}+\Delta u^{\mu}, 
        \\ \pi^{\mu \nu}=\pi_{0}^{\mu \nu}+\Delta \pi^{\mu \nu},\qquad B^{\mu}=B_{0} b_{0}^{\mu}+b_{0}^{\mu} \Delta B+B_{0} \Delta b^{\mu}
    \end{split}
\end{equation}
where we assume the following unperturbed shear-stress tensor
\begin{equation}
\label{eq: parallel shear}
  \pi_{0}^{\mu \nu}=\pi_{\|}(\tau)\left(b_{0}^{\mu} b_{0}^{\nu}+\frac{\Xi^{\mu \nu}}{2}\right).
\end{equation}
Here, $\pi_\parallel(\tau)$ denotes  the longitudinal component of shear stress tensor along the magnetic field and varies in time. The projection tensor,  $\displaystyle{\Xi^{\mu\nu} = \Delta^{\mu\nu} + b^\mu b^\nu}$ represents the projection onto the space orthogonal to both the velocity and the magnetic field.
We now restrict our analyses to perturbations \textit{parallel to the magnetic field}.  In the limit of asymptotically large magnetic fields, the transverse components of the shear stress tensor simplifies considerably and its transverse components become proportional to the longitudinal component of the shear-stress tensor. In Fourier space, one finds the following simple relation:
\begin{equation}
\Delta \tilde{\pi}_{bk}  \simeq \frac{3}{2}\pi_\parallel \Delta\tilde{b}_k,\qquad
\Delta \tilde{\pi}_{bq}  \simeq \frac{3}{2}\pi_\parallel \Delta\tilde{b}_q,    
\end{equation}
where the tilde denotes the Fourier transform of the corresponding field and the indices $b$, $k$, $q$ denote, respectively, the direction along the magnetic field, along the perturbation wave-vector, and the direction orthogonal to both -- the magnetic field and the perturbations. Thus,  $\Delta \tilde{\pi}_{bk}$  and $\Delta \tilde{\pi}_{bq}$ represent the transverse perturbations of the shear stress tensor  in Fourier space while $\Delta\tilde{b}_k$ and $\Delta\tilde{b}_q$ represent the transverse components of the perturbation of the magnetic field in the Fourier space.  
Substituting these asymptotic values into the conservation laws and combining them with the linearized Maxwell equations, we obtain
\begin{align}
    \left(\epsilon+P+B_0^2-\frac{\pi_\parallel}{2}\right)\partial_T^2\Delta\tilde{u}_k + \left(B_0^2 -\frac{3}{2}\pi_\parallel\right) k_b^2\Delta\tilde{u}_k &= 0, \\
    \left(\epsilon+P+B_0^2-\frac{\pi_\parallel}{2}\right)\partial_T^2\Delta\tilde{u}_q + \left(B_0^2 -\frac{3}{2}\pi_\parallel\right) k_b^2\Delta\tilde{u}_q &= 0, 
\end{align}
Here, $\Delta\tilde{u}_k$ ($\Delta \tilde{u}_q$) represents the perturbation in four-velocity in the $k$ direction (and in the $q$ direction) in Fourier space. Both equations lead to the same dispersion relation, 
\begin{equation}
    \omega^2 =\frac{B_0^2 -\frac{3}{2}\pi_\parallel}{\epsilon+P+B_0^2-\frac{\pi_\parallel}{2}} k_b^2
\end{equation}
which reduces to the usual Alfv\'en mode when $\pi_\| = 0$.   A physical instability arises when $\omega^2<0$, i.e., 
\begin{equation}
    B_0^2 - \frac{3}{2}\pi_\| <0.
\end{equation}
This is a well-known inequality, referred to as the \textit{Firehose Instability}, where the anisotropic stress (or parallel pressure) exceeds the magnetic pressure. This reduces the effective magnetic tension thus causing the plasma to decouple from the magnetic field and whip apart. In this regime, field lines become unstable and perturbations grow exponentially. This emphasizes that the perpendicular components of the shear stress-tensor induced by $\pi_\|$,  are not necessarily small when this instability sets in.

\section{Conclusion}
We derived the equations of relativistic magnetohydrodynamics for the dilute plasmas present in accretion disks, in the presence of strong magnetic fields. We modeled the system as a binary mixture of ions and electrons within the framework of the Boltzmann-Vlasov equation. The resulting MHD equations for the shear stress tensor cannot be captured by an Israel-Stewart-like formalism due to intrinsic coupling between the total shear stress tensor, $\pi^{\mu\nu}$ and the relative shear stress tensor, $\delta\pi^{\mu\nu}$.
Employing the approximation of mass asymmetry, we analyzed the equations in the linear regime and found that in strongly magnetized accretion disks, when the longitudinal component is sufficiently large, our equations lead to the well known firehose instability, in which the anisotropic stress overcomes the magnetic pressure, reducing magnetic tension and driving exponential growth of the perturbations. In this regime, our novel MHD framework suggests that the transverse components of the shear stress tensor may be important for correctly describing the plasma's dynamics.

\bibliography{references}

@article{Denicol_2018,
   title={Nonresistive dissipative magnetohydrodynamics from the Boltzmann equation in the 14-moment approximation},
   volume={98},
   ISSN={2470-0029},
   url={http://dx.doi.org/10.1103/PhysRevD.98.076009},
   DOI={10.1103/physrevd.98.076009},
   number={7},
   journal={Physical Review D},
   publisher={American Physical Society (APS)},
   author={Denicol, Gabriel S. and Huang, Xu-Guang and Molnár, Etele and Monteiro, Gustavo M. and Niemi, Harri and Noronha, Jorge and Rischke, Dirk H. and Wang, Qun},
   year={2018},
   month=oct }

@article{Denicol_resistive,
  title = {Resistive dissipative magnetohydrodynamics from the Boltzmann-Vlasov equation},
  author = {Denicol, Gabriel S. and Moln\'ar, Etele and Niemi, Harri and Rischke, Dirk H.},
  journal = {Phys. Rev. D},
  volume = {99},
  issue = {5},
  pages = {056017},
  numpages = {11},
  year = {2019},
  month = {Mar},
  publisher = {American Physical Society},
  doi = {10.1103/PhysRevD.99.056017},
  url = {https://link.aps.org/doi/10.1103/PhysRevD.99.056017}
}

@article{Hattori_2022,
   title={New Developments in Relativistic Magnetohydrodynamics},
   volume={14},
   ISSN={2073-8994},
   url={http://dx.doi.org/10.3390/sym14091851},
   DOI={10.3390/sym14091851},
   number={9},
   journal={Symmetry},
   publisher={MDPI AG},
   author={Hattori, Koichi and Hongo, Masaru and Huang, Xu-Guang},
   year={2022},
   month=sep, pages={1851} }

@article{Armas_2022,
   title={A stable and causal model of magnetohydrodynamics},
   volume={2022},
   ISSN={1475-7516},
   url={http://dx.doi.org/10.1088/1475-7516/2022/10/039},
   DOI={10.1088/1475-7516/2022/10/039},
   number={10},
   journal={Journal of Cosmology and Astroparticle Physics},
   publisher={IOP Publishing},
   author={Armas, Jay and Camilloni, Filippo},
   year={2022},
   month=oct, pages={039} }

@article{DNMR,
   title={Derivation of transient relativistic fluid dynamics from the Boltzmann equation},
   volume={85},
   ISSN={1550-2368},
   url={http://dx.doi.org/10.1103/PhysRevD.85.114047},
   DOI={10.1103/physrevd.85.114047},
   number={11},
   journal={Physical Review D},
   publisher={American Physical Society (APS)},
   author={Denicol, G. S. and Niemi, H. and Molnár, E. and Rischke, D. H.},
   year={2012},
   month=jun }

@article{Chandra_2015,
   title={AN EXTENDED MAGNETOHYDRODYNAMICS MODEL FOR RELATIVISTIC WEAKLY COLLISIONAL PLASMAS},
   volume={810},
   ISSN={1538-4357},
   url={http://dx.doi.org/10.1088/0004-637X/810/2/162},
   DOI={10.1088/0004-637x/810/2/162},
   number={2},
   journal={The Astrophysical Journal},
   publisher={American Astronomical Society},
   author={Chandra, Mani and Gammie, Charles F. and Foucart, Francois and Quataert, Eliot},
   year={2015},
   month=sep, pages={162} }

@article{Foucart_2015,
   title={Evolution of accretion discs around a kerr black hole using extended magnetohydrodynamics},
   volume={456},
   ISSN={1365-2966},
   url={http://dx.doi.org/10.1093/mnras/stv2687},
   DOI={10.1093/mnras/stv2687},
   number={2},
   journal={Monthly Notices of the Royal Astronomical Society},
   publisher={Oxford University Press (OUP)},
   author={Foucart, Francois and Chandra, Mani and Gammie, Charles F. and Quataert, Eliot},
   year={2015},
   month=dec, pages={1332–1345} }

@article{Most_2022,
   title={Modelling general-relativistic plasmas with collisionless moments and dissipative two-fluid magnetohydrodynamics},
   volume={514},
   ISSN={1365-2966},
   url={http://dx.doi.org/10.1093/mnras/stac1435},
   DOI={10.1093/mnras/stac1435},
   number={4},
   journal={Monthly Notices of the Royal Astronomical Society},
   publisher={Oxford University Press (OUP)},
   author={Most, Elias R and Noronha, Jorge and Philippov, Alexander A},
   year={2022},
   month=may, pages={4989–5003} }

@article{Cordeiro_2024,
   title={Causality Bounds on Dissipative General-Relativistic Magnetohydrodynamics},
   volume={133},
   ISSN={1079-7114},
   url={http://dx.doi.org/10.1103/PhysRevLett.133.091401},
   DOI={10.1103/physrevlett.133.091401},
   number={9},
   journal={Physical Review Letters},
   publisher={American Physical Society (APS)},
   author={Cordeiro, Ian and Speranza, Enrico and Ingles, Kevin and Bemfica, Fábio S. and Noronha, Jorge},
   year={2024},
   month=aug }

@article{Kushwah_2024,
   title={Relativistic dissipative magnetohydrodynamics from the Boltzmann equation for a two-component gas},
   volume={109},
   ISSN={2470-0029},
   url={http://dx.doi.org/10.1103/PhysRevD.109.096021},
   DOI={10.1103/physrevd.109.096021},
   number={9},
   journal={Physical Review D},
   publisher={American Physical Society (APS)},
   author={Kushwah, Khwahish and Denicol, Gabriel S.},
   year={2024},
   month=may }
\end{document}